\documentclass[preprint,showpacs,preprintnumbers,amsmath,amssymb,superscriptaddress]{revtex4-1}

\usepackage[dvips]{graphicx}
\usepackage{amsmath,bm}
\usepackage{comment}

\usepackage{color}

\begin{document}


\title{Period Variability of Coupled Noisy Oscillators}


\author{Fumito Mori}
\email[]{mori.fumito@ocha.ac.jp}
\affiliation{Department of Information Sciences, Ochanomizu University, Tokyo, Japan}
\author{Hiroshi Kori}%
\affiliation{Department of Information Sciences, Ochanomizu University, Tokyo, Japan}
\affiliation{PRESTO, Japan Science and Technology Agency, Kawaguchi, Japan}


\date{\today}

\begin{abstract}
Period variability, quantified by the standard deviation (SD) of the
cycle-to-cycle period, is investigated for noisy phase oscillators.
We define the checkpoint phase as the beginning/end point of one
oscillation cycle and derive an expression for the SD as a function
of this phase. We find that the SD is dependent on the checkpoint
phase only when oscillators are coupled. The applicability of our
theory is verified using a realistic model.
Our work clarifies the
relationship between period variability and synchronization from
which valuable information regarding coupling can be inferred.

\end{abstract}


\maketitle


Oscillators functioning as clocks, such as crystal oscillators
\cite{zhou08}, spin torque oscillators
\cite{rippard05,kaka05,mancoff05,keller09}, and circadian and heart
pacemakers \cite{winfree01, reppert02, glass01}, play an important
role in various systems. Although these clocks are subjected to
various types of noise, including thermal, quantum, and molecular
noise, they are required to perform temporally precise oscillations;
i.e., oscillations with only a small variability in the period
(known as ``period jitter'' in electronic engineering
\cite{yamaguchi01jitter}).

In many cases, it is sufficient for the clock to strike precisely at
a specific time in each oscillation cycle, and thus a perfectly
regular oscillation waveform is not needed.  For cardiac pacemakers
only the moment of stimulation is relevant.  Experimental data
regarding circadian activity in mice \cite{herzog04} indicate that
the variability in the period between each activity onset is smaller
than that between each offset. Similar results have also been
obtained in explant circadian pacemaker tissue (the suprachiasmatic
nucleus, SCN) \cite{herzog04}.  These observations suggest that the onset is more
important than the offset in
a circadian clock,
which may be designed in such a way that the crucial moment is expressed with
high precision.



\begin{figure}[h]
\begin{center}
\includegraphics[width=8cm]{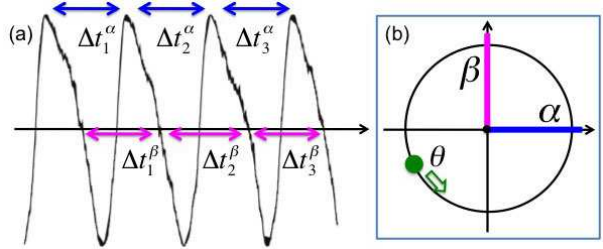}
\end{center}
\caption{(color online). (a) An example of the time series of an
oscillation. Periods are observed at two checkpoints, $\alpha$ and
$\beta$. (b) The corresponding checkpoint phases in the phase
description.
}
\label{fig:fig1}
\end{figure}

Remember that the definition of an oscillation period requires a
fixed beginning/end point for each oscillation cycle; hereafter
referred to as the checkpoint (Fig. \ref{fig:fig1}). Although the
average period does not depend on the particular choice of
checkpoint, the period variability may be sensitive to the
checkpoint. In order to clarify whether the checkpoint dependence in
circadian activity is an artifact due to a technical problem in
determining the onset and offset times or an essential property of
the circadian clock, we need to investigate under what conditions
the period variability is dependent on the checkpoint; this has
received scant attention to date.


Another important aspect of the period variability is its
relationship to synchronization. A clock is commonly synchronized to
its master clock such as in the case of the SCN in response to the
daily variation of sunlight, and in peripheral clocks in response to
the SCN. In addition, most biological clocks, including the SCN,
cardiac pacemakers, and pacemakers in weakly electrical fish, are
composed of a population of synchronized oscillators
\cite{winfree01,reppert02, moortgat00}. It is known, both
experimentally and theoretically, that period variability is reduced
when the oscillators are coupled and synchronized \cite{clay79,
enright80, winfree01, needleman01,kojima06, kori11}. The question,
therefore, arises as to whether the checkpoint dependence of the
period variability is attributable to the interaction between
oscillators.

In this Letter, we discuss this checkpoint dependence for the case
of coupled noisy phase oscillators.
The period variability can be quantified using the standard
deviation (SD) of the cycle-to-cycle period, and we show that
although the SD is not dependent on the checkpoint in a single phase
oscillator, it is dependent in a system of coupled phase
oscillators; i.e., the checkpoint dependence results from the
coupling effect.
The SD is derived as a function of the checkpoint phase, which
clarifies the relationship between the SD and synchronization.
In particular, we find that in the case of diffusive coupling
between oscillators, the checkpoint dependence of the SD has the
same tendency as that of the synchronization: the SD is small when
the oscillators are well synchronized.  In other cases, however, the
relationship is more complex. We also apply our theory to a
realistic model of the electrical activity in a cell to demonstrate
its validity. We believe that this is the first theoretical study to
elucidate the existence of precise timing and its relationship with
synchronization.

To begin, we prove that the period variability is independent of the
checkpoint in a single phase oscillator system. When a limit cycle
oscillator is subjected to weak noise, its dynamics are well
described by the following phase oscillator model
\cite{winfree67,kuramoto84};
\begin{equation}
\frac{d \theta}{dt}= \omega + Z(\theta)  \sqrt{D} \xi (t),
\label{eq:single}
\end{equation}
where $\theta$ and  $\omega$ are the phase and natural frequency,
respectively. The $2\pi$-periodic function $Z(\theta)$ is a phase
sensitivity function, which quantifies the phase response of the
oscillator to noise, and $ \xi(t)$ denotes independent and
identically distributed (i.i.d.)  noise;
each random variable $\xi(t)$ for all $t$ obeys the same probability
distribution
and all are mutually independent.
The positive constant $D$ denotes the noise strength.
Note that our proof below holds even if we permit $\omega$ and the
probability distribution of $\xi$ to be 2$\pi$-periodic functions of
$\theta$:  $\omega(\theta)$ and $\xi(t,\theta)$.

The $k$th oscillation time of an oscillator,
$t_{k}^{\theta_{\text{cp}}}$, is defined as the time at which
$\theta$ passes through 2$\pi k + \theta_{\text{cp}}$ $(0  \le
\theta_{\text{cp}} < 2\pi)$ for the first time [Fig.
\ref{fig:fig1}(b)]. We define $\theta_{\text{cp}}$ as the checkpoint
phase. The $k$th oscillation period $\Delta
t_{k}^{\theta_{\text{cp}}}$ is defined as $\Delta
t_{k}^{\theta_{\text{cp}}}= t_{k}^{\theta_{\text{cp}}}-
t_{k-1}^{\theta_{\text{cp}}}$, and the SD is defined as
\begin{equation}
{\text{SD}}
(\theta_{\text{cp}})=\sqrt{E[({\Delta t_{k}^{\theta_{\text{cp}}}}-\tau)^{2}]},
\label{eq:defsd}
\end{equation}
where $E[\cdots]$ represents the statistical average over $k$, and
$\tau$ is the average period given by $\tau=E[\Delta
t_{k}^{\theta_{\text{cp}}}]$. Note that $E[\cdots]$ denotes both the
statistical average taken over $k$ and the ensemble average in the
present paper, which are identical in the steady state. The system
given by Eq.~(\ref{eq:single})
 is always in the steady state.

To prove that the SD is independent of $\theta_{\text{cp}}$, we
introduce two checkpoint phases denoted by ${\alpha}$ and ${\beta}$
[Fig. 1(b)]. Since  the processes $\alpha \rightarrow \beta$ and
$\beta \rightarrow \alpha$ for any $k$ are independent, we arrive at
${\text{SD}}(\alpha)={\text{SD}}(\beta)$ for any arbitrary
checkpoint phases $\alpha$ and $\beta$.
A detailed proof is given in 
Appendix A.

Next, we consider a pair of coupled phase oscillators subjected to
noise. When limit cycle oscillators are weakly coupled to each other
and subjected to weak noise, the dynamics can be described by
\cite{winfree67, kuramoto84}
\begin{eqnarray}
\left \{
\begin{array}{l}
  \dot{{\theta_{1}}}
= \omega
+ \kappa J(\theta_{1},\theta_{2})+  Z(\theta_{1}) \sqrt{D}    \xi_{1}
(t), \\
  \dot{{\theta_{2}}}
 = \omega
+ \kappa J(\theta_{2},\theta_{1})+     Z(\theta_{2})  \sqrt{D}  \xi_{2} (t) ,
\end{array}
\right .
\label{eq:coupled}
\end{eqnarray}
where $\theta_{i}$ and $\kappa \ge 0$ are the phase of the
oscillator $i$ and the coupling strength, respectively. The i.i.d.
noise $\xi_{i} (t)$
satisfies $E[\xi_{i}(t)]=0$ and $E[\xi_{i}(t)
\xi_{j}(t')]=\delta_{ij}\delta(t-t')$.
The $2\pi$-periodic function $J(x,y)$ describes the interaction
between oscillators, which leads to synchronization.
We assume that, in the absence of noise ($D=0$), the oscillators are
synchronized in phase, i.e., $\theta_{1,2}(t) \rightarrow \phi (t) $
$(t\rightarrow \infty) $, where $\phi (t)$  is a solution of
\begin{equation}
\dot{\phi} (t)=\omega+ \kappa J(\phi,\phi).
\label{eq:phi}
\end{equation}
The necessary condition for the stability of in-phase synchrony for
$D=0$ is provided below [see Eq.~(\ref{eq:nc})]. We also assume that
$\omega+\kappa J(\phi,\phi)>0$ for any $\phi$
for the coupled system to be oscillatory.


Our particular interest is in the relationship between the SD [Eq.
(\ref{eq:defsd})] and the synchronization of two oscillators. We
thus introduce the following order parameter that measures the phase
distance from the in-phase state:
\begin{equation}
d  (\theta_{\text{cp}})  =\sqrt{
 E \left[  \| \theta_{1} - \theta_{2} \|^{2}         \right]
_{\theta_{1} = \theta_{\text{cp}}}   },
\label{eq:defd}
\end{equation}
where $E[x(t)]_{\theta_{1}= \theta_{\text{cp}}}$ represents the
average of $x_k$ over $k$ (where $x_k$ is the value of $x(t)$ taken
when $\theta_{1}$ passes through $2\pi k + \theta_{\text{cp}}$ for
the first time), and $\| \theta_{1} - \theta_{2} \|$ is the phase
difference defined on the ring $[-\pi ,\pi)$.
The phase distance $d(\theta_{\text{cp}})$
is zero when the oscillators are completely synchronized in phase,
and increases with
 the phase difference.

As we demonstrate below, the relationship between
SD$(\theta_{\text{cp}})$ and $d(\theta_{\text{cp}})$ is
qualitatively different for the two cases where $J(\phi,\phi)$ is
(A) independent of $\phi$
and (B) dependent on $\phi$.
{C}ases (A) and (B) imply that
$\dot \phi$ given in Eq.~(\ref{eq:phi})
is independent of $\phi$ and dependent on $\phi$, respectively.
Phase reduction theory indicates that it is appropriate to assume
the form $J(x,y)=z(x)G(x,y)$, where $z(x)$ is the phase sensitivity
function for the interaction $G(x,y)$
\cite{winfree67,kuramoto84}. It is known that diffusive coupling
between chemical oscillators and gap-junction coupling between cells
yields $J(x,y)=z(x)(h(x)-h(y))$, where $h$ represents a chemical
concentration \cite{kori01, kiss05} or membrane potential,
which corresponds to case (A).
%
Case (A) also allows the form $J(x,y)=j(x-y)$, which has been
employed in many models such as the Kuramoto model
\cite{kuramoto84};
however, we
do not employ
 this form
in the demonstration,
since the term $j(x-y)$ is derived as a result of averaging the
interaction $z(x)G(x,y)$ over one oscillation period
\cite{kuramoto84}, and, by this approximation, the information about
the $\theta_{\text{cp}}$ dependence is lost. Many other types of
coupling, such as $J(x,y)=z(x)h(y)$ employed below, correspond to
case (B) \cite{ariaratnam01}.



As an example of case (A), we consider $z(\theta)=\sin{\theta}$ for
$0 \le \theta < \pi$, $z(\theta)=0$ for $\pi  \le \theta < 2\pi$,
and $h(\theta)=\cos{\theta}$, and the following as an example of
case (B): $z(\theta)=-\sin{\theta}$ and $h(\theta)=1+\cos{\theta}$ \cite{ariaratnam01}.
We set $Z(\theta)=1$,  $\omega=2\pi$, $\sqrt{D}=0.03 \times 2 \pi$,
and $\theta_{1}(0)=\theta_{2}(0)=0$, and assume $\xi_{1,2}(t)$ to be
white Gaussian noise. We integrate Eq. (\ref{eq:coupled}) using the
{Euler} scheme with a time step of $5 \times 10^{-4}$ for
$t=0$--$10100$ and discard the $t=0$--$100$ data as transient.

\begin{figure}[h]
\begin{center}
\includegraphics[width=8.5cm]{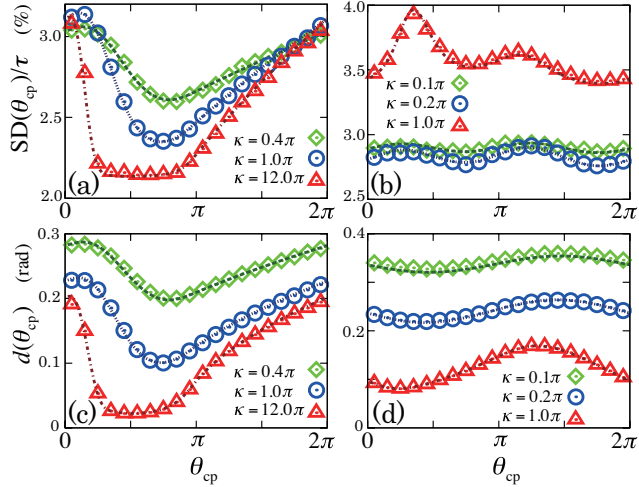}
\end{center}
\caption{(color online). The SD$(\theta_{\text{cp}})/\tau$
for case (A) and (B) is shown in (a) and (b), respectively,
where the vertical scale is
expressed as a percentage.
The distance
from in-phase
synchronization,  $d(\theta_{\text{cp}})$,
for case (A) and (B) is shown in (c) and (d), respectively.
The points and lines are the
numerical results of the simulation and analytical predictions
given by Eqs. (\ref{eq:SD}) and (\ref{eq:dsyn}), respectively.  }
\label{fig:CVA}
\end{figure}

Using these examples, numerically obtained SD values for
$\theta_{1}$ are plotted as a function of $\theta_{\text{cp}}$ in
Fig. \ref{fig:CVA}(a) and (b).
{The results} indicate clearly the existence of $\theta_{\text{cp}}$
dependence
in both cases, which was absent in the single phase oscillator
system.
%
%
This dependence becomes stronger for larger $\kappa$ values.
In contrast, for $\kappa \ll \omega$,
the dependence vanishes
because $J(x,y)$ is well approximated by $j(x-y)$ \cite{kuramoto84},
and thus, the system effectively has rotational symmetry.
{The $\theta_{\text{cp}}$ value at which
SD$(\theta_{\text{cp}})$ assumes its minimum} represents the most
precise timing.

The $\theta_{\text{cp}}$ dependence of $d{(\theta_{\text{cp}})}$ for
the two cases is shown in Fig. \ref{fig:CVA} (c) and (d).
A comparison with SD$(\theta_{\text{cp}})$ shows that the checkpoint
phase maxima and minima of each $\kappa$ value coincide in the case of
(A). Thus, the most precise timing is obtained when the oscillators
are synchronized. By contrast, the $\theta_{\text{cp}}$ dependence
is
considerably different in the case of (B).
Therefore, we expect that nontrivial factors, apart from
synchronization, influence the SD.
We also examined several other functions, $z(\theta)$, $h(\theta)$,
and $Z(\theta)$, and found a similar relationship between
$\text{SD}(\theta_{\text{cp}})$ and $d(\theta_{\text{cp}})$ (data
not shown).

We now derive an expression for the SD. The derivation consists of
two steps:
(i) calculation of
the phase diffusion $\sigma (\theta_{\text{cp}})$ [defined by
Eq.~(\ref{eq:sigma})]
with a linear approximation, and (ii) transformation from
$\sigma (\theta_{\text{cp}})$ to SD($\theta_{\text{cp}}$).
%
Here, we employ the solution $\phi(t)$ of Eq.~(\ref{eq:phi}) with $\phi(0)=0$
and
 the time $t_{\text{cp}}$ is defined by
$\phi(t_{\text{cp}})=\theta_{\text{cp}}$.
 The oscillation period
for $D=0$ is denoted by $\tau$; i.e., $\phi( t_{\text{cp}} +\tau)=
\theta_{\text{cp}}+2\pi$.
After a transient time, our system approaches the steady state,
which is defined by the following equation for all $\Psi$:
\begin{equation}
P(\|  \theta_{1}-\theta_{2} \|; \theta_{1}=\Psi)=P(\|  \theta_{1}-\theta_{2} \|  ; \theta_{1}=\Psi+2\pi),
\label{eq:steady}
\end{equation}
where $P(\|  \theta_{1}-\theta_{2} \|; \theta_{1}=\Psi) $ is the
probability density function of the distance $\|
\theta_{1}-\theta_{2} \|$
at $\theta_{1}= \Psi$.
%
We assume that the system is in the steady state at $t=0$.
The ensemble we consider here is defined by the initial condition at
$t=t_{\text{cp}}$, $\theta_{1}(t_{\text{cp}})=\theta_{\text{cp}}$,
and $\theta_{2}(t_{\text{cp}}) $ is distributed in
 $[\theta_{1}(t_{\text{cp}})-\pi, \theta_{1}(t_{\text{cp}}) + \pi)$ according to Eq.~(\ref{eq:steady}).
From this point, $E[\cdots]$ represents the average taken over this
ensemble.
The phase diffusion $\sigma (\theta_{\text{cp}})$ is defined by
\begin{equation}
 \sigma (\theta_{\text{cp}})^2 =
E[ (\theta_{1}(t_{\text{cp}}+\tau) -\theta_{1}(t_{\text{cp}})-2 \pi)^{2} ].
\label{eq:sigma}
\end{equation}
%
We also assume that the noise intensity $D$ is sufficiently small
and that the other parameters and functions are of $O(1)$,
so that the phase difference $\| \theta_1 - \theta_2 \|$ is small in
most cases in the steady state.

To calculate the phase diffusion, we decompose $\theta_{1,2} $ as
$\theta_{1,2} (t) =\phi(t) + \Delta_{1,2}(t)$.
We then consider the time duration $0\le t \le O(\tau)$, in which
$\Delta_{1,2}(t) \ll 1$ is expected in most cases because $D\ll 1$.
Therefore, we can linearize Eq.~(\ref{eq:coupled}).
%
We define the two modes, $X=\Delta_{1}+\Delta_{2}$ and
$Y=\Delta_{1}-\Delta_{2}$, which obey
\begin{equation}
(\dot{X}, \dot{Y})
=\kappa
f_{X,Y} (\phi(t)) (X,Y)
+
\xi_{X,Y} (t,\phi(t)),
\label{eq:XYdot}
\end{equation}
where $f_{X} (\phi) \equiv
  \left.  \frac{\partial J}{\partial x} \right|_{x=y=\phi}
+   \left.     \frac{\partial J}{\partial y} \right|_{x=y=\phi }
 =\frac{d J(\phi,\phi)}{d \phi}$,
$f_{Y} (\phi) \equiv  \left.  \frac{\partial J}{\partial x}
\right|_{x=y=\phi} -   \left.     \frac{\partial J}{\partial y}
\right|_{x=y=\phi } $,
and $\xi_{X,Y}(t,\phi(t)) \equiv  \sqrt{D}
Z(\phi(t)) (\xi_{1}(t) \pm \xi_{2}(t))$.
Note that
$f_{X} (\phi)=0$ for all $\phi$ in case (A).
The solutions
of Eq. (\ref{eq:XYdot}) can be described as
\begin{eqnarray}
 && ({X},{Y})(t)=\exp{ \left[+\kappa F_{X,Y}(\phi(t)) \right] }     \nonumber \\
 && \times     \left\{   (X,Y)(0)
+ \int_{0}^{t} \exp{  \left[ - \kappa F_{X,Y}(\phi(t')) \right] } \xi_{X,Y}(t',\phi(t'))  dt'                 \right\} ,
\end{eqnarray}
where
$F_{X,Y}(\phi(t)) \equiv \int_{0}^{t}  f_{X,Y}(\phi(t')) dt'$.
Furthermore, because $f_{X} (\phi) =\frac{d J(\phi,\phi)}{d \phi}$,
we obtain
\begin{equation}
 F_{X}(\phi(t))= \frac{1}{\kappa} \ln \left( \frac{\dot{\phi} (t)}{\dot{\phi} (0)} \right).
\label{eq:FX}
\end{equation}
%
For $\xi_{Y}=0$, we obtain
$F_{Y}(2\pi)=(1/\kappa)\ln(Y(\tau)/Y(0))$.
Therefore, in the absence of noise, in-phase synchronization is
stable if
\begin{equation}
F_{Y}(2 \pi)  \equiv c <0.
\label{eq:nc}
\end{equation}

The correlations,  $E[X(t)^{2}]$, $E[Y(t)^{2}]$, and $E[X(t)Y(t)]$
are given in Appendix B. 
Since Eq.~(\ref{eq:steady}) can be rewritten as
$P(| \Delta_{1}-\Delta_{2} |; t ) \cong P( |  \Delta_{1}-\Delta_{2} |  ;
t+\tau)$, then $ E[Y(t)^{2}]= E[Y(t+\tau)^{2}]$ holds approximately,
leading to
\begin{equation}
E[Y(0)^{2}]=2D \frac{\exp[2 \kappa c ]} { 1-\exp[2 \kappa c ]}
 \int_{0}^{\tau}   Z(\phi(t'))^{2} \exp[-2 \kappa F_{Y}(\phi(t'))] dt'.
\end{equation}
%
In addition, because $d(\theta_{\text{cp}})^{2} = E[
(\Delta_{1}(t_{\text{cp}}) - \Delta_{2}(t_{\text{cp}}))^{2}] =
E[Y(t_{\text{cp}})^{2}]$, we obtain
\begin{eqnarray}
{d(\theta_{\text{cp}})}^{2}    &=& \exp[2\kappa F_{Y}(\theta_{\text{cp}})]    \nonumber \\
&&\times  \left(  E[Y(0)^{2}]  +  2D \int_{0}^{\theta_{\text{cp}}}
Z(\phi)^{2}
  \exp[-2\kappa F_{Y}(\phi)] \frac{ds}{d\phi (s)}  d \phi  \right),
\label{eq:dsyn}
\end{eqnarray}
which is generally $\theta_{\text{cp}}$-dependent even if $Z(\phi)$ is constant.

Using these correlations and  Eq.~(\ref{eq:FX}),
%
%
we obtain the following expression for the phase diffusion 
(See Appendix B) 
\begin{eqnarray}
 \sigma (\theta_{\text{cp}})^2
   & = & E[(\Delta_{1}(t_{\text{cp}}+\tau) -\Delta_{1}(t_{\text{cp}}) )^{2}]     \nonumber \\
& =&C_{1} {\dot{\phi}(\theta_{\text{cp}})}^{2}+C_{2}
 {d}(\theta_{\text{cp}})^{2},
\label{eq:pdif}
\end{eqnarray}
where the $C_{1,2}$ are independent of $\theta_{\text{cp}}$ and are
given by
$C_{1}=\frac{D}{2} \int_{0}^{2 \pi}    \frac{Z(\theta)^{2}}{\dot{\phi}(\theta)^{3}}  d \theta $
and
$C_{2}=(1-\exp[\kappa c  ])/2$.
The $C_{1}$ term is an effective diffusion constant for the center
of the two oscillators, which is half that of an uncoupled
oscillator, and the $C_{2}$ term is associated with the stability of
the synchronization.


To transform $\sigma (\theta_{\text{cp}})$ to
SD$(\theta_{\text{cp}})$,
we note that when the noise intensity is low, most of
the trajectories of $\theta_{1}(t)$ are
very close
to the
unperturbed trajectory $\phi(t)$
(see Appendix C).
 In such a case, the following relation approximately
holds true:
\begin{equation}
\frac{ \sigma (\theta_{\text{cp}})}
{{\text{SD}}(\theta_{\text{cp}})}
= \dot{\phi}(\theta_{\text{cp}}).
\label{eq:trans}
\end{equation}
The same approximation (but for constant $\dot \phi$) was employed in
Ref. \cite{kori11} and verified numerically.

From Eqs. (\ref{eq:pdif}) and (\ref{eq:trans}), we finally arrive at
\begin{equation}
{\text{SD}}(\theta_{\text{cp}})
=\sqrt{C_{1}+C_{2} \frac{ {d(\theta_{\text{cp}})}^{2} }{ {\dot{\phi}(\theta_{\text{cp}})}^{2} }}.
\label{eq:SD}
\end{equation}
The analytical results given by Eqs.
(\ref{eq:SD}) and (\ref{eq:dsyn}) are in excellent agreement with
the numerical results (Fig. \ref{fig:CVA}).
%
Although we have only discussed paired identical phase oscillators,
our theory can easily be extended to other cases, e.g., $N$ globally
coupled (all-to-all) identical oscillators or a periodically driven
noisy oscillator.%

Equation (\ref{eq:SD}) shows that the periodicity of
SD($\theta_{\text{cp}}$) is based on
the synchronization $d(\theta_{\text{cp}})$
and  phase velocity $\dot{\phi}(\theta_{\text{cp}})$.
For case (A), since $\dot{\phi}(\theta_{\text{cp}})$ is constant,
there is one-to-one correspondence between SD$(\theta_{\text{cp}})$ and
$d(\theta_{\text{cp}})$; i.e.,
the most precise timing ($\theta_{\text{cp}}^{\text{min}}$) is the
timing at which the best synchronization is achieved. This was
observed in Fig. \ref{fig:CVA} (a) and (c), where
$\theta_{\text{cp}}^{\text{min}}=\pi/2+O(\kappa^{-1}) $ can be
obtained from $ d   d(\theta_{\text{cp}}) /d \theta_{\text{cp}}=0 $.
For case (B), however, the SD also depends on
$\dot{\phi}(\theta_{\text{cp}})$; this is in contrast to that
observed for the single phase oscillator system in which the phase
velocity 
$\omega(\theta)$ does not contribute to the checkpoint dependence of
the SD. Figures \ref{fig:CVA}(b) and (d) showed that SD$(\theta_{\text{cp}})$
and $d(\theta_{\text{cp}})$ are considerably different, which
indicates the strong effect of $\dot \phi$ in this particular
example. Indeed, SD$(\theta_{\text{cp}})$ assumes its minimum around
a maximum $\dot \phi{(\theta_\text{cp})}$
($\theta_{\text{cp}}^{\text{min}} \approx 5\pi/3$).

\begin{figure}[h]
\begin{center}
\includegraphics[width=7cm]{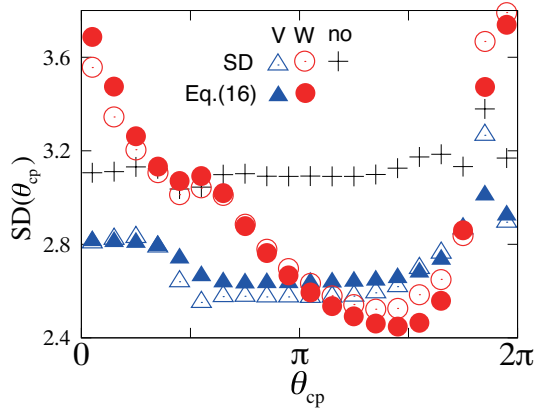}
\end{center}
\caption{(color online). Validation of Eq.~(\ref{eq:SD}) in  the
FitzHugh-Nagumo model.
Open symbols are the numerically obtained SD values. Filled symbols
are the SD values evaluated from Eq.~(\ref{eq:SD}) with the
numerically obtained $d$ values and fitting parameters $C_1$ and
$C_2$.
The triangles and circles represent the $V$- and $W$-coupling cases.
The plus symbols are the numerically obtained SD values for an
uncoupled oscillator.
}
\label{fig:fit}
\end{figure}

To investigate whether Eq.~(\ref{eq:SD}) holds for a more realistic
model, we employ the FitzHugh-Nagumo model given by
\begin{eqnarray}
\left \{
\begin{array}{l}
\dot{V_{1}}= V_{1}(V_{1}-a)(1-V_{1})-W_{1} +\xi_{1}(t) +K_{V} (V_{2}-V_{1}), \\
\dot{W_{1}} =  \epsilon (V_{1}-bW_{1}) +K_{W} (W_{2}-W_{1}),
\end{array}
\right .
\label{eq:FN}
\end{eqnarray}
in which the second oscillator is described in a similar way. We
fixed $a=-0.1$, $b=0.5$, and $\epsilon=0.01$. This system
shows limit-cycle oscillations with a period of $\tau \simeq 126.5$
when noise and coupling are absent. The white Gaussian noise
$\xi_{i}(t)$ has an intensity of $0.01$. The interaction is
diffusive, i.e., case (A), and we consider the following two types:
$V$-coupling   $(K_{V}=0.01, K_{W}=0)$ and $W$-coupling  $(K_{V}=0,
K_{W}=0.01)$. The phase $\theta$  was defined properly 
(see Appendix D),
and ${\text{SD}}(\theta_{\text{cp}})$ and ${d(\theta_{\text{cp}})}$
were obtained numerically. Figure \ref{fig:fit} shows that the
$\theta_{\text{cp}}$-dependence of the SD is different in the two
cases, suggesting a significant effect from the coupling. We
estimated the $C_1$ and $C_2$ values using Eq. (16) and the
least-squares method under the condition that both cases have the
same $C_{1}$ value, resulting in $C_{1}=5.4$, $C_{2}^{(V)}=0.20$,
and $C_{2}^{(W)}=0.48$.
In Fig. \ref{fig:fit}, we can see that the SD is described well by
Eq.~(\ref{eq:SD}) using the fitted $C_{1}$ and $C_{2}$ values.
%
%
This demonstrates that the theory is valid  for this biological model.

In many cases, only the SD measured at a functionally
relevant checkpoint characterizes the performance of a clock.
 When designing a precise clock,
we only have to reduce
SD$(\theta_{\text{cp}})$  for a specific $\theta_{\text{cp}}$.
 Equation (\ref{eq:SD}) implies that
SD$(\theta_{\text{cp}})$ at a given $\theta_{\text{cp}}$ decreases
with decreasing $d(\theta_{\text{cp}})$ and increasing $\dot
\phi(\theta_{\text{cp}})$. Therefore, attractive coupling between
oscillators should be activated around the functionally relevant
timing point.
In addition, in case (B), the phase velocity should be increased
through coupling.
%

Our theory enables us to infer the coupling timing or form by measuring
SD$(\theta_{\rm cp})$ at several checkpoints.
Although this is, in principle, possible with $d(\theta_{\rm cp})$,
using SD$(\theta_{\rm cp})$ has the added advantages 
that the SD can be measured from a single time series and that
$d(\theta_{\rm cp})$ is sensitive to the definition of phase.
From the observations of circadian periods in mice described in the
introduction \cite{herzog04}, it is possible that the SCN sends
signals to the peripheral clocks around the onset of a subjective
day.
%
An experimental observation of the checkpoint dependence in other
biological clocks
would be a new source of coupling information.

We thank Hiroshi Ito for valuable discussions.
This work was supported by JSPS KAKENHI Grant Number 23$\cdot$11148.

\appendix

\section{Proof that the SD is independent of the checkpoint phase in a single phase oscillator}
We introduce two checkpoint phases denoted by ${\alpha}$ and
${\beta}$. By defining the intervals $\Delta t_{k}^{\beta
\rightarrow \alpha}=t_{k}^{{\alpha}}- t_{k}^{{\beta}}$ and $\Delta
t_{k}^{\alpha \rightarrow \beta}=t_{k}^{{\beta}}- t_{k-1}^{\alpha}$,
the oscillation periods  observed at $\alpha$ and $\beta$ can be
decomposed as $\Delta t_{k}^{\alpha}=\Delta t_{k}^{\beta \rightarrow
\alpha}+\Delta t_{k}^{\alpha \rightarrow \beta}$ and $\Delta
t_{k}^{\beta}=\Delta t_{k}^{\alpha \rightarrow \beta}+\Delta
t_{k-1}^{\beta \rightarrow \alpha}$, respectively. Because $\xi(t)$
is independent, the processes $\alpha \rightarrow \beta$ and $\beta
\rightarrow \alpha$ for any $k$ are independent. We thus have
\begin{equation}
E[\Delta t_{k}^{\beta \rightarrow \alpha} ]=E[\Delta t_{k-1}^{\beta \rightarrow \alpha}],
\end{equation}
\begin{equation}
E[(\Delta t_{k}^{\beta \rightarrow \alpha})^{2} ]=E[(\Delta
t_{k-1}^{\beta \rightarrow \alpha})^{2}],
\end{equation}
and
\begin{equation}
 E[  \Delta t_{k}^{\alpha \rightarrow \beta} \Delta t_{k-1}^{\beta \rightarrow \alpha} ]
=E[  \Delta t_{k}^{\alpha \rightarrow \beta}]E[ \Delta t_{k-1}^{\beta \rightarrow \alpha} ]
  =   E[  \Delta t_{k}^{\alpha \rightarrow \beta} \Delta t_{k}^{\beta \rightarrow \alpha} ].
\end{equation}
The average and the mean square period are independent of the 
checkpoint phase 
labels; i.e.,
\begin{equation}
E[\Delta t_{k}^{\alpha}]=E[\Delta t_{k}^{\beta \rightarrow \alpha}]+E[\Delta t_{k}^{\alpha \rightarrow \beta}]
=E[\Delta t_{k-1}^{\beta \rightarrow \alpha}]+E[\Delta t_{k}^{\alpha \rightarrow \beta}]
=E[\Delta t_{k}^{\beta}]=\tau
\end{equation}
and
\begin{equation}
E[(\Delta t_{k}^{\alpha})^{2}]=E[(\Delta t_{k}^{\beta \rightarrow \alpha})^{2}]+E[(\Delta t_{k}^{\alpha \rightarrow \beta})^{2}]
+2E[\Delta t_{k}^{\beta \rightarrow \alpha}][ \Delta t_{k}^{\alpha \rightarrow \beta}]
=E[(\Delta t_{k}^{\beta})^{2}].
\end{equation}
Thus, we arrive at
\begin{equation}
{\text{SD}}(\alpha)= \sqrt{E[({\Delta t_{k}^{\alpha}}-\tau)^{2}]}
= \sqrt{E[({\Delta t_{k}^{\beta}}-\tau)^{2}]}
={\text{SD}}(\beta)
\end{equation}
for any arbitrary checkpoint phases $\alpha$ and $\beta$.

\section{Calculation of the correlations}
The correlations of the noise terms, $\xi_{X,Y}(t,\phi(t)) =
\sqrt{D}Z(\phi(t)) (\xi_{1}(t) \pm \xi_{2}(t))$, are given as
\begin{eqnarray}
E[ \xi_{X}(s,\phi(s))  \xi_{X}(s',\phi(s'))]
  & =  & 2 D Z(\phi(s)) Z(\phi(s'))  \delta(s-s'),   \label{eq:noixx}  \\
  E[ \xi_{Y}(s,\phi(s))  \xi_{Y}(s',\phi(s'))]
  & =  & 2 D Z(\phi(s)) Z(\phi(s'))  \delta(s-s'),     \label{eq:noiyy} \\
    E[ \xi_{X}(s,\phi(s))  \xi_{Y}(s',\phi(s'))]
  & =  & 0.   \label{eq:noixy}
\end{eqnarray}
%
  Using
  Eqs.~(9), (\ref{eq:noixx}), (\ref{eq:noiyy}), (\ref{eq:noixy}), and $E[\xi_{X,Y}(t,\phi(t))]=0$,
we obtain
\begin{eqnarray}
 E[X(t)^{2}]  &=&  \exp{[ 2 \kappa F_{X} (\phi(t))   ]}
  \biggl[
E[X(0)^{2}]   \nonumber   \\
&& +  \int_{0}^{t}  \int_{0}^{t}   \exp[-\kappa \{F_{X}(\phi (s))+ F_{X}(\phi (s'))\}]   E[ \xi_{X}(s,\phi(s))  \xi_{X}(s',\phi(s'))]     ds ds'
\biggr]       \nonumber  \\
& =& \exp{[ 2 \kappa F_{X} (\phi(t))   ]}
  \biggl[
E[X(0)^{2}]
 +2D \int_{0}^{t}  Z(\phi(s))^{2}      \exp[-2\kappa F_{X}(\phi (s))] ds  \biggr],  \label{eq:xx}  \\
 E[Y(t)^{2}]  &=&  \exp{[ 2 \kappa F_{Y} (\phi(t))   ]}
  \biggl[
E[Y(0)^{2}]   \nonumber   \\
&& +  \int_{0}^{t}  \int_{0}^{t}   \exp[-\kappa \{F_{Y}(\phi (s))+ F_{Y}(\phi (s'))\}]   E[ \xi_{Y}(s,\phi(s))  \xi_{Y}(s',\phi(s'))]     ds ds'
\biggr]       \nonumber  \\
& =& \exp{[ 2 \kappa F_{Y} (\phi(t))   ]}
  \biggl[
E[Y(0)^{2}]
 +2D \int_{0}^{t}  Z(\phi(s))^{2}      \exp[-2\kappa F_{Y}(\phi (s))] ds  \biggr],
\end{eqnarray}
and
\begin{eqnarray}
E[X(t)Y(t)] &  = & \exp{[  \kappa  ( F_{X} (\phi(t))    +  F_{Y} (\phi(t)) )  ]}
E[X(0)Y(0)]. \label{eq:xy}
\end{eqnarray}
Substituting $t=t_{\text{cp}}+\tau$ in
 Eqs.~(\ref{eq:xx}) and (\ref{eq:xy}),
we obtain
\begin{eqnarray}
E[X(t_{\text{cp}} +\tau)^{2}]
&=&
\exp{[ 2 \kappa F_{X} ( \theta_{\text{cp}} + 2 \pi )   ]}
  \biggl[ E[X(0)^{2}]
 +2D \int_{0}^{t_{\text{cp}}+\tau}  Z(\phi(s))^{2}      \exp[-2\kappa F_{X}(\phi (s))] ds  \biggr] \nonumber \\
 &=&
  \exp{[ 2 \kappa F_{X} ( \theta_{\text{cp}} )]}
   \biggl[ E[X(0)^{2}]
   +2D \int_{0}^{t_{\text{cp}}}  + \int_{t_{\text{cp}}}^{t_{\text{cp}}+\tau}
     Z(\phi(s))^{2}      \exp[-2\kappa F_{X}(\phi (s))] ds  \biggr]    \nonumber \\
     & = &
     E[X(t_{\text{cp}} )^{2}]
     + 2D \exp[2 \kappa F_{X}(\theta_{\text{cp}})]
\int_{0}^{\tau}
Z(\phi(s))^{2}  \exp[-2\kappa F_{X}(\phi(s))] ds,
\label{eq:xxtau}
\end{eqnarray}
and
\begin{eqnarray}
E[X(t_{\text{cp}} +\tau) Y(t_{\text{cp}} +\tau)]
&=& \exp{[  \kappa  ( F_{X} ( \theta_{\text{cp}}+2 \pi )    +  F_{Y} ( \theta_{\text{cp}}+2 \pi ) )  ]}   E[X(0)Y(0)]   \nonumber \\
&  = &  \exp{[ \kappa c  ]}  E[X(t_{\text{cp}} )  Y(t_{\text{cp}} )  ],
\label{eq:xytau}
\end{eqnarray}
where we use $\phi(t_{\text{cp}}+\tau)=\theta_{\text{cp}}+2 \pi$,
$F_{X}(\theta+2 \pi)=F_{X}(\theta)$, $F_{Y}(\theta+2 \pi)=F_{Y}(\theta) +c$, and
$Z(\theta+2 \pi)=Z(\theta)$.
Inserting $\theta_{1}(t_{\text{cp}})=\theta_{\text{cp}}$ and $\phi(t_{\text{cp}})=\theta_{\text{cp}}$
into the definition $\Delta_{1}(t)=\theta_{1}(t)-\phi(t)$,
we obtain
 \begin{equation}
 \Delta_{1}(t_{\text{cp}})=0.
\end{equation}
We then obtain
\begin{equation}
E[X(t_{\text{cp}})^{2}]
=-E[X(t_{\text{cp}}) Y(t_{\text{cp}}) ] =E[Y(t_{\text{cp}})^{2}]
=d(\theta_{\text{cp}})^{2}.
\label{eq:d2}
\end{equation}

Using Eqs.~(\ref{eq:xxtau})--(\ref{eq:d2}), the relation
$E[Y(t_{\text{cp}} +\tau)^{2}]=E[Y(t_{\text{cp}} )^{2}]$,
 and Eq.(10),
we obtain the following expression for the phase diffusion
\begin{eqnarray}
 \sigma (\theta_{\text{cp}})^2
   & = & E[(\Delta_{1}(t_{\text{cp}}+\tau) -\Delta_{1}(t_{\text{cp}}) )^{2}]     \nonumber \\
&=& \frac{1}{4} \left\{    E[X(t_{\text{cp}}+\tau)^{2}] +E[Y(t_{\text{cp}}+\tau)^{2}] +2E[X(t_{\text{cp}}+\tau) Y(t_{\text{cp}}+\tau)]       \right\}      \nonumber   \\
&=& \frac{1}{4} \left\{    2D \exp[2 \kappa F_{X}(\theta_{\text{cp}})]
\int_{0}^{\tau}
Z(\phi(t'))^{2}  \exp[-2\kappa F_{X}(\phi(t'))] dt'
+2(1- \exp{[ \kappa c  ]} ) {d}(\theta_{\text{cp}})^{2}
    \right\}      \nonumber   \\
& =&C_{1} {\dot{\phi}(\theta_{\text{cp}})}^{2}+C_{2}
 {d}(\theta_{\text{cp}})^{2}.
\label{eq:pdif}
\end{eqnarray}

\newpage

\section{Transformation from phase diffusion to period variability}
\begin{figure}[h]
\begin{center}
\includegraphics[width=16cm]{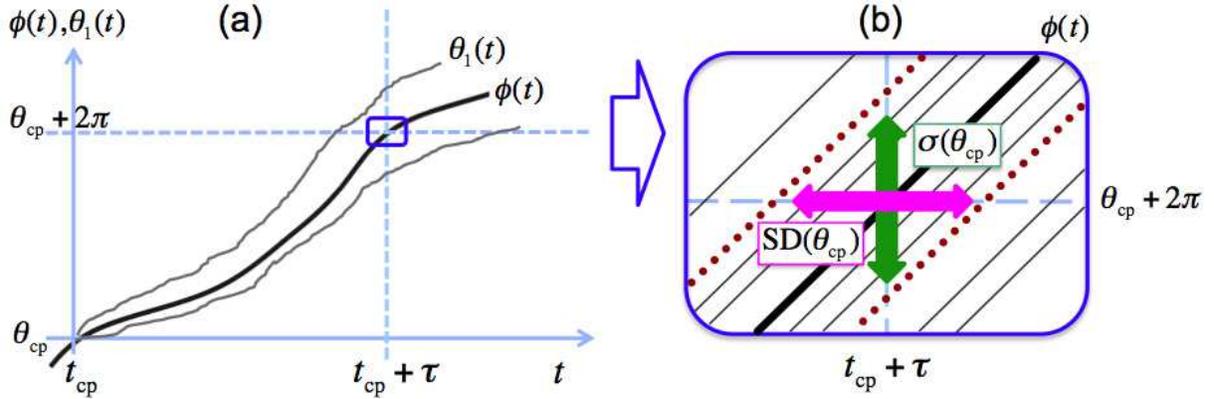}
\end{center}
\caption{Illustration of the relationship between $\sigma
(\theta_{\text{cp}})$ and ${{\text{SD}}(\theta_{\text{cp}})}$.
 }
\label{fig:trans}
\end{figure}

Here, we illustrate the relationship between $\sigma
(\theta_{\text{cp}})$ and ${{\text{SD}}(\theta_{\text{cp}})}$.
Figure~\ref{fig:trans}(a) presents a schematic view of the
trajectories of $\phi(t)$ and $\theta_{1}(t)$.
An enlarged view of the region around $(t_{\text{cp}}+\tau,
\theta_{\text{cp}}+2 \pi)$
is displayed in Fig.~\ref{fig:trans}(b), in which the vertical width
between the dotted lines represents the standard deviation of the
phase distribution of $\theta_{1}(t)$. In particular, the vertical
arrow represents the standard deviation of
$\theta_{1}(t_{\text{cp}}+\tau)$, which is denoted by $\sigma
(\theta_{\text{cp}})$.
Because we assume a low noise intensity, the actual trajectories of
$\theta_{1}(t)$ (thin lines) are very close to that of $\phi(t)$.
We can thus expect that the trajectories are 
approximately straight 
and 
 parallel to $\phi(t)$ 
in this enlarged region. Therefore,
the horizontal width between the dotted lines at
$\theta_{1}=\theta_{\text{cp}}+2 \pi$ is approximately equal to
${{\text{SD}}(\theta_{\text{cp}})}$ (horizontal arrow), and the
relation $\sigma
(\theta_{\text{cp}})/{{\text{SD}}(\theta_{\text{cp}})}=\dot{\phi}(\theta_{\text{cp}})$
holds approximately.

\section{Definition of phase in the FitzHugh-Nagumo model}
\begin{figure}[h]
\begin{center}
\includegraphics[width=8cm]{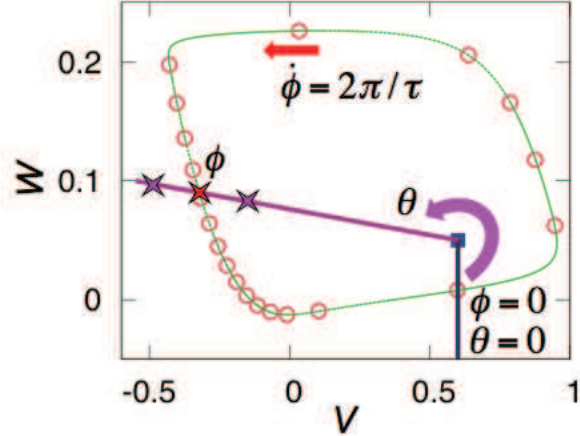}
\end{center}
\caption{Illustration of the definition of the phase in the
FitzHugh-Nagumo model. The limit cycle trajectory is generated by a
coupled FitzHugh-Nagumo model without noise, whose parameters are
given in the main text. The circles are placed at equally spaced
intervals of $\phi$. All points on a straight line radiating from
the origin (filled square) have the same phase.} \label{fig:phase}
\end{figure}

We define the  phase $\theta$ as a function of $(V, W)$, which are
the state variables of the FitzHugh-Nagumo model given in  Eq.~(17)
in the main text, as follows (Fig.~\ref{fig:phase}).
We first assign $\phi$ values to all points on the limit cycle
trajectory generated by Eq.~(17)
without noise such that $\phi$ identically satisfies $\dot {\phi}=2
\pi/ \tau$, where $\tau$ is the period. The limit cycle trajectory
is independent of the coupling strengths.
We set $\phi=0$ at $V=0.6$ with $\dot{V}>0$.
We then consider radial lines extending from an arbitrary point
inside the limit cycle, which we chose as $(0.6, 0.05)$ (filled
square) in this case. When a radial line intersects the limit cycle
at a point that has a value of $\phi$,
the phase $\theta$ of all points on the radial line is defined by  $\theta=\phi$.
These radial lines are different from isochrones that give a
standard definition of the phase \cite{kuramoto84}, 
but 
the isochrones 
are usually unknown. As
shown in Fig. 3, our theory is valid even for this practical
definition. 

%

\end{document}